\documentclass{aastex631}

\graphicspath{{./}{figures/}}

\begin{document}

\title{Reaction kinetics of CN + toluene and its implication on the productions of aromatic nitriles in the Taurus molecular cloud and Titan’s atmosphere}

\author[0000-0001-9764-2576]{Mengqi Wu}
\affiliation{Hefei National Laboratory for Physical Sciences at the Microscale, University of Science and Technology of China, Hefei, Anhui, 230026, P. R. China.}

\author[0000-0001-8651-7833]{Xiaoqing Wu}
\affiliation{Hefei National Laboratory for Physical Sciences at the Microscale, University of Science and Technology of China, Hefei, Anhui, 230026, P. R. China.}
\affiliation{College of Information Engineering, China Jiliang University, Hangzhou 310018, P. R. China.}

\author[0000-0002-9999-7298]{Qifeng Hou}
\affiliation{Hefei National Laboratory for Physical Sciences at the Microscale, University of Science and Technology of China, Hefei, Anhui, 230026, P. R. China.}

\author[0000-0002-3153-3534]{Jiangbin Huang}
\affiliation{Hefei National Laboratory for Physical Sciences at the Microscale, University of Science and Technology of China, Hefei, Anhui, 230026, P. R. China.}

\author[0000-0002-8212-9435]{Dongfeng Zhao}
\affiliation{Hefei National Laboratory for Physical Sciences at the Microscale, University of Science and Technology of China, Hefei, Anhui, 230026, P. R. China.}

\author[0000-0002-9730-8487]{Feng Zhang}
\affiliation{Hefei National Laboratory for Physical Sciences at the Microscale, University of Science and Technology of China, Hefei, Anhui, 230026, P. R. China.}

\correspondingauthor{Dongfeng Zhao}
\email{dzhao@ustc.edu.cn}

\correspondingauthor{Feng Zhang}
\email{feng2011@ustc.edu.cn}



\begin{abstract}
Reactions between cyano radical and aromatic hydrocarbons are believed to be important pathways for the formation of aromatic nitriles in the interstellar medium (ISM) including those identified in the Taurus molecular cloud (TMC-1). Aromatic nitriles might participate in the formation of polycyclic aromatic nitrogen containing hydrocarbons (PANHs) in Titan’s atmosphere. Here, ab initio kinetics simulations reveal a high efficiency of $\rm \sim10^{-10}~cm^{3}~s^{-1}$ and the competition of the different products of 30–-1800 K and $10^{-7}$--100 atm of the CN + toluene reaction. In the star-forming region of TMC-1 environment, the product yields of benzonitrile and tolunitriles for CN reacting with toluene may be approximately 17$\%$ and 83$\%$, respectively. The detection of main products, tolunitriles, can serve as proxies for the undetected toluene in the ISM due to their much larger dipole moments. The competition between bimolecular and unimolecular products is extremely intense under the warmer and denser PANH forming region of Titan’s stratosphere. The computational results show that the fractions of tolunitriles, adducts, and benzonitrile are 19$\%$--68$\%$, 15$\%$--64$\%$ and 17$\%$, respectively, at 150–-200 K and 0.0001–-0.001 atm (Titan’s stratosphere). Then, benzonitrile and tolunitriles may contribute to the formation of PANHs by consecutive $\rm C_{2}H$ additions. Kinetic information of aromatic nitriles for the CN + toluene reaction calculated here helps to explain the formation mechanism of polycyclic aromatic hydrocarbons (PAHs) or PANHs under different interstellar environments and constrains corresponding astrochemical models.

\end{abstract}


\section{Introduction} \label{sec:intro}

Polycyclic aromatic hydrocarbons (PAHs) and their derivatives, presumably, the nitrogen containing analogues (PANHs), are important constituents of the interstellar medium (ISM) \citep{1,59,60,61}. Based on the ubiquitously observed infrared emission bands in space, these species are inferred to contribute 10--$\rm 25\%$ of the total cosmic carbon \citep{57,58,62}, and play key roles in the evolution of interstellar biological systems. They are also present in planetary atmospheres, such as that of Saturn's largest moon, Titan, where PA(N)Hs has been derived from mass spectrometric and spectroscopic data \citep{14,15,63} and are likely important prebiotic molecules and precursors of aerosols in the brownish-red organic haze layers of Titan’s atmosphere. However, identification of specific PAHs by radio astronomy is always challenging, because many of them have relatively small rotational constants for these large molecules, and for pure PAHs very small permanent dipole moments \citep{3,4}. Very recently, a number of small aromatic molecules, including a number of aromatic nitriles (benzonitrile, cyanonaphtalene, and cyanoindene), have been discovered in the prototypical cold dark Taurus molecular cloud TMC-1 via rotational transitions, and are considered to be precursor molecules in the formation of larger PANHs \citep{5,6,7}.

The CN radical, although highly reactive, is known to be ubiquitously and abundantly present in the ISM, circumstellar shells, and planetary atmospheres \citep{10,11}. Its reactions with small aromatic hydrocarbons may play an important role in the formation mechanism of PANHs in the ISM \citep{9,12,13}. E.g., in TMC-1, the detected benzonitrile, cyanonaphtalene and cyanoindene, are suggested to be formed by CN reactions with benzene, naphthalene and indene, respectively, in the currently available astrochemical models. However, predicted column densities of the detected aromatic nitrilesby the models are found to be underestimated \citep{5,6,7,8}. Therefore, in-depth investigations for CN radical reactions with aromatic hydrocarbons helps understand the formation and evolution of PA(N)Hs in the interstellar medium and in Titan’s atmosphere.

The focus of the present study is on the reaction kinetics of toluene + CN under the temperature-pressure conditions relevant to both cold, harsh interstellar environments and upper atmosphere of Titan. This reaction also serves as a model system for understanding the CN reactions with methyl-substituted aromatic molecules. Joblin et al. \citep{2} concluded that methyl- and ethylene-substituted PAHs may be potential candidates responsible for the weak 3.4 $\rm \mu m$ emission feature of observed infrared emission bands. Recently Cernicharo et al. \citep{16} reported a high 3$\sigma$ upper limit of $\rm 6~\times~10^{12}~cm^{-2}$ for toluene in the QUIJOTE1 line survey. Their model shows that large uncertainty of the toluene's abundance brings large uncertainty to the models of the TMC-1. In Titan’s upper atmosphere, toluene has been detected by the Cassini ion and neutral mass spectrometer \citep{17}, and Loison et al.'s model also shows that toluene is one of the most abundant substituted benzene molecules there \citep{18}. Few experimental and theoretical investigations have focused on the reactions of CN radical with toluene, which may be a potential candidate contributing to the formation of benzonitrile in the ISM \citep{19}. Previous experimental results on the CN + toluene reaction are from Trevitt et al. \citep{20} and Messinger et al. \citep{19}. Trevitt et al. \citep{20} first used a pulsed Laval nozzle expansion coupled with laser-induced fluorescence detection to measure the rate coefficients of the CN + toluene reaction and obtained only a value of $\rm (1.3~\pm~0.3)~\times~10^{-10}~cm^{3}~s^{-1}$ at 105 K due to the nonexponential decays of CN at room temperature in the presence of toluene. They also discussed product detections from their synchrotron VUV multiplexed photoionization mass spectrometry (MPIMS) measurements. The only detected product is the cyanotoluene constitutes with no significant detection of benzyl radicals. Messinger et al. \citep{19} reported rate coefficients of $\rm (4.1~\pm~0.2)~\times~10^{-10}~cm^{3}~s^{-1}$ over the temperature range of 15--294 K, as measured by a Cin\'{e}tique de R\'{e}action en Ecoulement Supersonique Uniforme (CRESU) apparatus coupled with the pulsed laser photolysis-laser-induced fluorescence (PLP-LIF) technique. They concluded that there is negligible pressure and temperature dependence for the total rate coefficients under the studied conditions. The authors also theoretically calculated the potential energy surface (PES) with M06-2X/aug-cc-pVTZ method, which is the only theoretical result for the CN + toluene reaction. Their theoretical results revealed that the barrierless addition-dissociation pathways with exit barriers are well below the energy of the reactant show great advantages under low temperature conditions. In contrast to the consistent experimental results of the CN + benzene reaction with rate coefficients of $\rm \sim4~\times~10^{-10}~cm^{3}~s^{-1}$ \citep{20,21,22}, the discrepancy between the two studies on the CN + toluene reaction reached a factor of 3. The lack of product yields of CN +toluene reaction raises questions on the structure effect of the aromatic molecule on its reaction mechanism for forming CN-substitute compounds, which motivates us to study it from the view of kinetic theories.

Here, high-level quantum chemical calculations combined with RRKM/ME simulations was used to explore the reaction mechanics and chemical kinetics of the CN + toluene reaction under conditions of 30--1800 K and $10^{-7}$--100 atm, covering the conditions related to molecular clouds, planetary atmosphere, and circumstellar environments \citep{23}. The chosen Rice-Ramsperger-Kassel-Marcus/Master equation (RRKM/ME) method has been used to obtain the temperature- and pressure-dependent product yields for the reaction systems of CN radical with various molecules including methylamine and ethylene \citep{13,24,25,26}. Based on the computed PES and rate coefficients, the feasibility of identifications of the related products for the titled reaction in the ISM is then discussed. Kinetic information obtained by theoretical calculations here provides new insights into the the tholins’ formation in Titan’s atmosphere and constains the model of TMC-1.

\section{Methodology} \label{sec:Methodology}

\begin{figure}[htbp]
\centering
\includegraphics[width=0.5\linewidth]{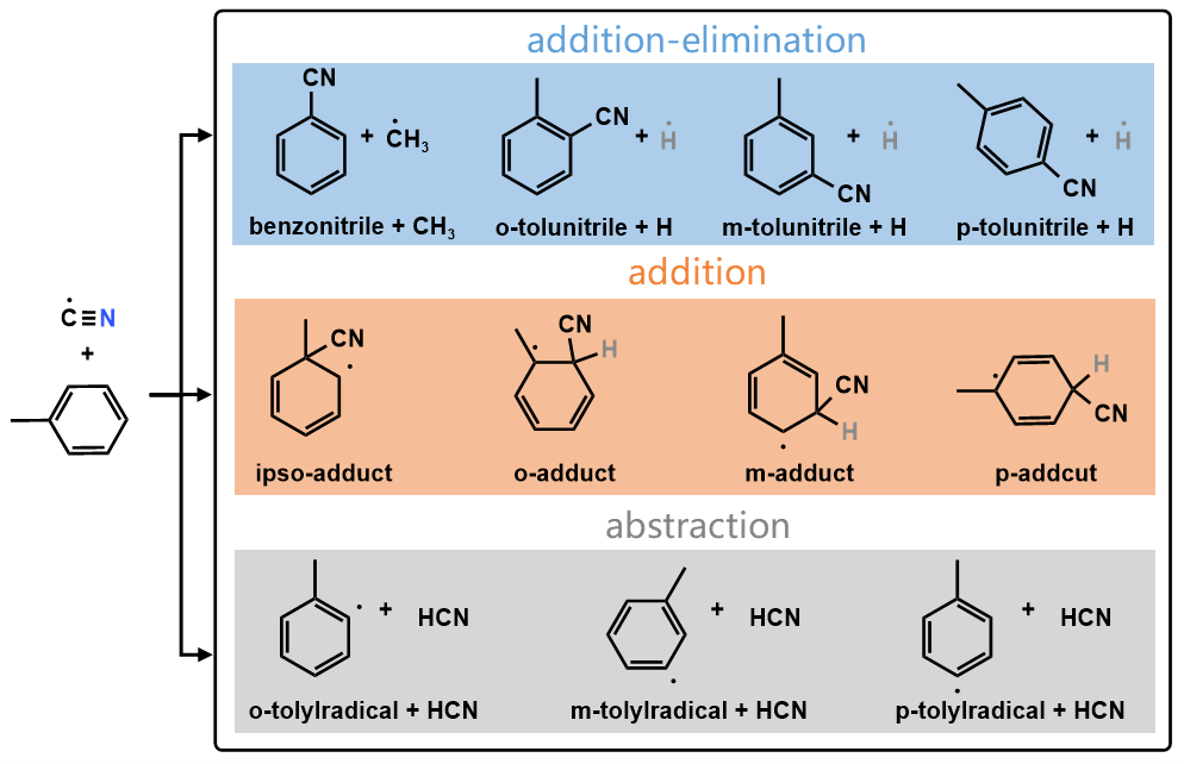}
\caption{Main reaction types and corresponding products for the CN + toluene reaction.  \label{fig:f1}}
\end{figure}

The hybrid-meta-exchange correlation functional, M06-2X \citep{27}, with D3 dispersion correction \citep{28} and 6-311+G(d,p) basis sets \citep{29,30,31} was chosen to obtain the geometries of the reactants, products, and transition states (TSs) on the PES of the CN + toluene reaction. The zero-point energies (ZPEs) and vibrational frequencies were computed at the same level with a scaling factor of 0.970 \citep{32}. The chosen M06-2X functional has shown excellent performance in studying main-group thermochemistry \citep{33}. Single point energy (SPE) calculations were implemented using the coupled-cluster single-, double-, and perturbative triple-excitations [CCSD(T)] method \citep{34}; the approximate complete basis set (CBS) limit  was evaluated with the help of the  Møller-Plesset second-order perturbation theory (MP2) calculations. The CCSD(T) energy is given by the sum of correlation energy and the SCF energy using the two-point extrapolation method via the following expression \citep{35}:
\begin{equation}
E(HL)=E_{SCF}(CBS)+E_{corr}(CCSD(T)/CBS)
\end{equation}
\begin{equation}
E_{corr}(CCSD(T)/CBS)=E_{corr}(MP2/CBS)+(E_{corr}(CCSD(T)/aVTZ)-E_{corr}(MP2/aVTZ))
\end{equation}
This calculation scheme has been applied in previous studies by Barone et al. and Baiano et al. \citep{24,36}. The T1 diagnostic values of stationary points and TSs suggest negligible multi-reference characteristics \citep{37,38}. Main reaction types and the corresponding products are shown in Figure \ref{fig:f1}. All M06-2X calculations were implemented with the Gaussian16 \citep{39} program, and the CCSD(T) calculations were performed with the ORCA 4.2.0 \citep{40,41} program package.

Based on the calculated results of PESs, the rate coefficients and branching ratios of various channels were computed by the RRKM/ ME method \citep{23}, with temperatures ranging from 30 to 1800 K and pressures ranging from $10^{-7}$ to 100 atm. We adopted the chemically significant eigenstate (CSE) approach \citep{42} to solve the one-dimensional master equation. Here, we chose $\rm N_{2}$ and $\rm H_{2}$ as the bath gases due to their abundant fractions in Titan’s atmosphere and TMC-1, respectively \citep{43}. The Lennard-Jones (L-J) potential model was applied to mimic the interactions between the C8H8N species ($\rm \sigma =5.96 $\AA, $ \epsilon =591.54K $) and the bath gas---$\rm N_{2}$ ($\rm \sigma =3.681 $\AA, $ \epsilon =67.89K $) and $\rm H_{2}$ ($\rm \sigma =2.920$\AA, $ \epsilon =26.41K $). The single-exponential down model was selected to describe the collisional energy transfer for this system with the average energy transferred in downward collisions of $\rm \left \langle \Delta E \right \rangle _{down} = 260\times (T/300) ^{0.875}~cm^{-1} $ analogously to the collisional parameters of $\rm N_{2}$ with toluene \citep{45}. The RRKM theory was adopted to obtain the microcanonical reactive flux except the barrierless addition. The one-dimensional hindered rotor (1D-HR) model was employed to treat the torsions of $\rm CH_{3}$ groups, and the Eckart tunneling correction \citep{46} was used for its simplicity. The phase space theory (PST) method \citep{47} assumes that the interaction between the two reacting fragments can be described by an isotropic potential, $\rm -C_{n}/R^{n}$. The PST method has been proven to be valid and efficient for the reactions of the CN radical with formaldehyde and acetaldehyde by Tonolo et al. under low-temperature conditions \citep{48}. Here, the exponential term was chosen to be 6, which was usually implemented in neutral-neutral reactions \citep{48}. The coefficient $\rm C_{6}$ was approximated by the sum of dipole-induced-dipole and dispersion terms \citep{49,50}. The finally applied $\rm C_{6}$ value was 387 Eh. Nevertheless, it is worth noting that the PST method may result in greater uncertainties when calculating the reaction flux of barrierless pathways at high temperatures. Despite this limitation, we chose to employ this method in our study due to its computational efficiency. The capture rate coefficients will be incorporated into the RRKM/ME simulation to obtain the final rate coefficients with a rough approximation that capture possibilities by each site are equal. All the RRKM/ME rate coefficients were computed with the MESS code \citep{23}.

\section{Results and discussion} \label{sec:discussion}

\subsection{Potential Energy Surface} \label{PES}

\begin{figure*}[bp]
\centering
\includegraphics[width=0.9\linewidth]{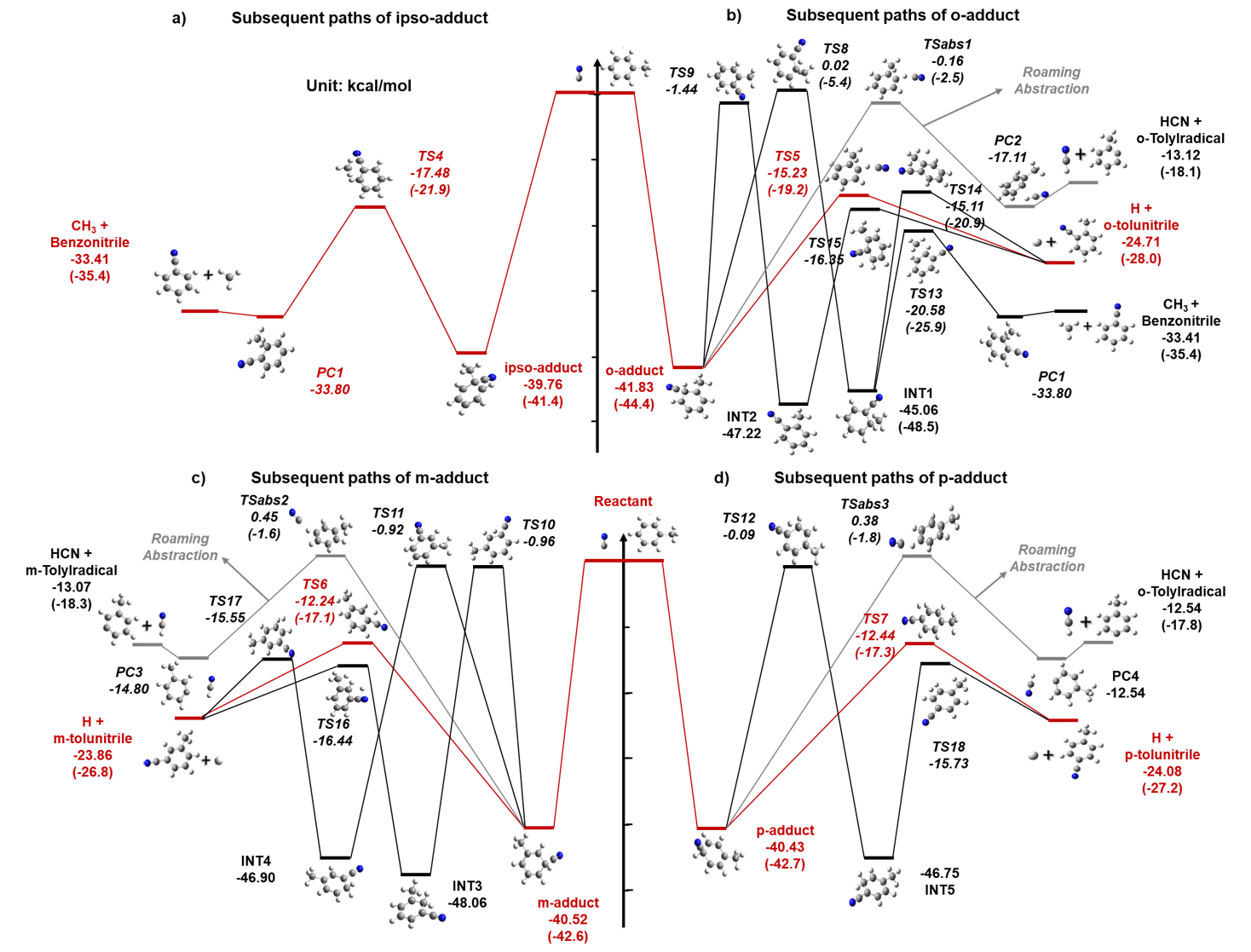}
\caption{Potential energy surface of the CN + toluene reaction at the level of CCSD(T)/CBS//M06–2X-D3/6-311+G(d,p) in this work. The energetically favorable pathways are represented in red. Energies in parentheses are the results calculated at the level of M06-2X/aug-cc-pVTZ from Messinger et al. \citep{19}.  \label{fig:f2}}
\end{figure*}

Figure \ref{fig:f2} illustrates the PES for the CN + toluene reaction at the level of CCSD(T)/CBS//M06-2X-D3/6-311+G(d,p), including four addition channels, four addition-elimination channels and three abstraction channels (Figure \ref{fig:f1}). Previous studies on the CN + benzene reaction have suggested that the isocyano products cannot compete with the cyano compounds in the ISM or Titan’s atmosphere \citep{9} due to the high barriers. Thus the pathways yielding isocyano compounds are not included here. 
The CN + toluene reaction begins with the C atom of the CN radical attacking the pi molecular orbital of toluene, which is consistent with the reaction between the CN radical and other unsaturated hydrocarbons \citep{13,49,51,52,53}. All of these addition processes are barrierless with exothermicities of nearly 40 kcal/mol, producing four adducts---ipso -adduct, o-adduct, m-adduct and p-adduct. There exist some barriers below the reactants between two different adducts and more details can be found in Figure \ref{fig:fA1} in the Appendix. The adducts can subsequently undergo isomerization, elimination and abstraction processes. The most energetically favorable reaction pathway is the elimination of $\rm CH_{3}$ group forming benzonitrile + $\rm CH_{3}$ with a submerged barrier of 22.28 kcal/mol.The elimination of the hydrogen atom forming H + o-, m-, p-tolunitrile with barriers of 26.6, 28.28, and 27.99 kcal/mol can also be competitive with the $\rm CH_{3}$ + benzonitrile channel, because all of the barriers are below the bimolecular reactants. These adducts can also isomerize with the hydrogen atom at the addition site migrating to adjacent sites, while the much higher barriers of the isomerization indicate that they cannot compete with the addition-elimination channels. The HCN + o-, m-, p-tolylradical products can be formed via a roaming H-abstraction mechanism \citep{52}, which means that the CN group leaves the adducts, roams around, and finally abstracts an H atom from the addition site. The roaming H-abstraction pathways show poor competitiveness compared to direct elimination channels due to their higher barriers. We failed to locate the TSs of direct H-abstraction. The previous results about the reactions between CN and other unsaturated hydrocarbons showed slightly positive barriers for direct H-abstraction, indicating its relatively unimportant role in the low-temperature kinetics of the CN + toluene reaction \citep{52}.

Theoretical results obtained by Messinger et al. at the level of M062X/aug-cc-pVTZ are also shown in parentheses in Figure \ref{fig:f2} for comparison \citep{19}. They revealed barrierless addition channels and energetically favorable addition-elimination channels as well, while partial structures for isomerization and abstraction channels of this work were not included in Messinger et al.'s calculations. The largest deviation between the energies at the level of M06-2X/aug-cc-pVTZ and our calculations by the CCSD(T)/CBS//M06-2X-D3/6-311+G(d,p) method is 5.47 kcal/mol, which could introduce great errors in the kinetic results---especially at low temperatures.

\subsection{Rate Coefficients} \label{sec:totalrate}

The related experimental data are especially limited for the CN + toluene reaction, including the only measurement of $\rm 1.3~\times~10^{-10}~cm^{3}~s^{-1}$ at 105 K from Trevitt et al. \citep{20} and an average of $\rm 4.1~\times~10^{-10}~cm^{3}~s^{-1}$ using CRESU apparatus at 15--294 K from Messinger et al. \citep{19}. Based on the computed  high-level PES, temperature and pressure dependent rate coefficients of the CN + toluene reaction were then calculated using the RRKM/ME theory, with temperatures ranging from 30 to 1800 K and the pressures ranging from $10^{-7}$ atm to 100 atm. 
\begin{figure*}[bp]
\centering
\includegraphics[width=0.9\linewidth]{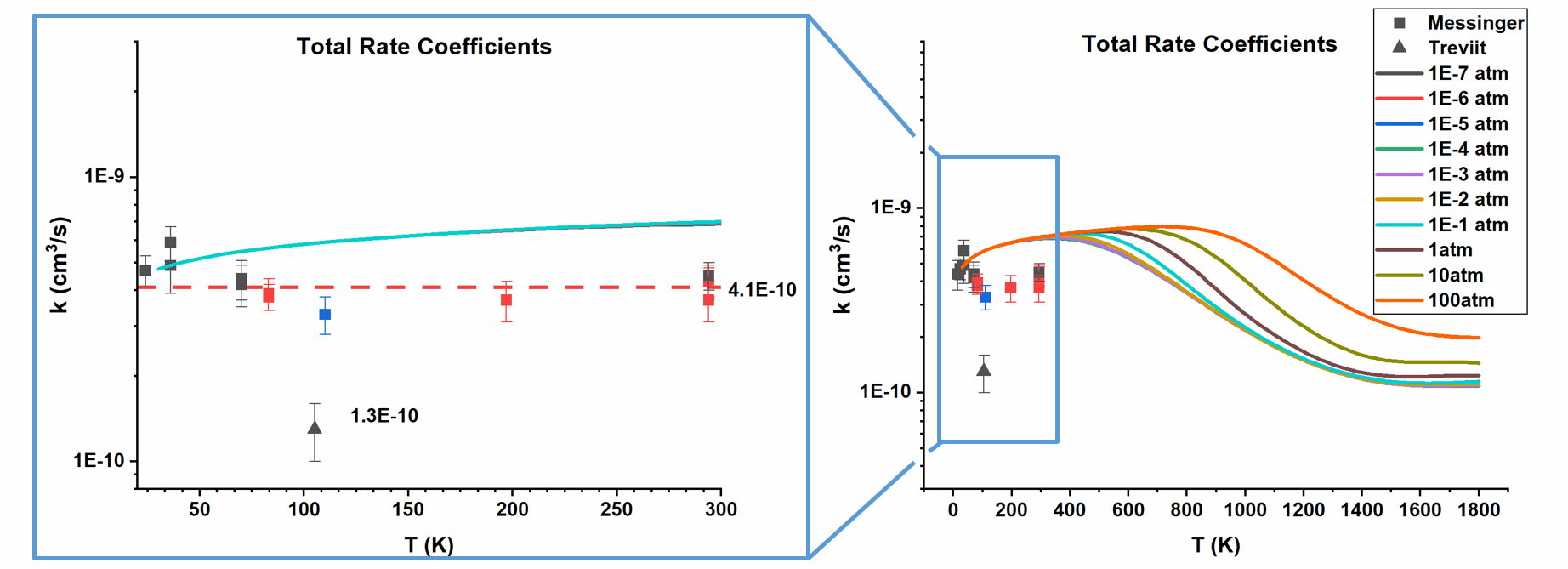}
\caption{Comparison of the total rate coefficients calculated in this work with previous experimental measurements for the CN + toluene reaction, with temperatures ranging from 30 to 1800 K and pressures ranging from $10^{-7}$ to 1000 atm. The contents inside the blue box are partially enlarged picture. Trevitt et al. \citep{20}, used a pulsed Laval nozzle expansion coupled with laser-induced fluorescence detection (total density: $\rm 2.4\times10^{16}~cm^{-3}$); Messinger et al. \citep{19}, used a CRESU apparatus coupled with the PLP-LIF technique (There were a series of pressures, and three types of bath gas: $\rm N_{2}$ shown by black square markers, He shown by red square markers, and Ar shown by blue square markers). The average value of the results from Messinger et al. is represented by a red dashed line ($\rm 4.1~\times~10^{-10}~cm^{3}~s^{-1}$).  \label{fig:f3}}
\end{figure*}
Figure \ref{fig:f3} shows the comparison of the calculated total rate coefficients in this work with previous experimental measurements. The calculated rate coefficients of the CN + toluene reaction approach the collisional limit with a range of $\rm 1.09~\times~10^{-10}$ to $\rm 7.96~\times~10^{-10}~cm^{3}~s^{-1}$, which is consistent with those of the reactions between CN and other unsaturated hydrocarbons \citep{13,49,51,52,53}. The total rate coefficients increase slightly with temperatures rising initially and then decrease sharply when the temperature goes over a critical value. From 100 to 200 K, there is generally a good agreement between the calculated rate coefficients in this work and the experimental measurements of Messinger et al. with a ratio ranging from 1.16 to 1.70. The corresponding deviation between this work and Messinger et al.’s results increases monotonously with temperature \citep{19}. However, our predictions overestimate the measurements of Trevitt et al. by a larger factor of 4.48 at 105 K. The calculations of energies and the collisional parameters show only a slight influence on the total rate coefficients, but these parameters have some impacts on branching ratios. Two different choices of bath gas obtained similar total rate coefficients. The rough approximation of the isotropic long-range potential used in the PST method fails to capture the chemical bonding interactions and short-range repulsions \citep{49}, which may contribute to a slight overestimation of the calculated rate coefficients at higher temperatures compared to experimental results. Nevertheless, considering the low temperatures of interest in this study, the uncertainties introduced by the PST method are deemed acceptable. For the high-temperature regime, more detailed theoretical methods such as ab initio molecular dynamics (AIMD), variable-reaction-coordinate variational transition state theory (VRC-VTST), or trajectory simulations are more suitable and precise. Unfortunately, the multidimensional PES calculations required by these methods are prohibitively expensive for the titled reaction and fall beyond the scope of this work. Rate coefficients for various channels with the bath gas of $\rm H_{2}$ can be also found in Table \ref{table:TA1} in the Appendix.

\subsection{Branching Ratios} \label{sec:Branching}

\begin{figure}[bp]
\centering
\includegraphics[width=1\linewidth]{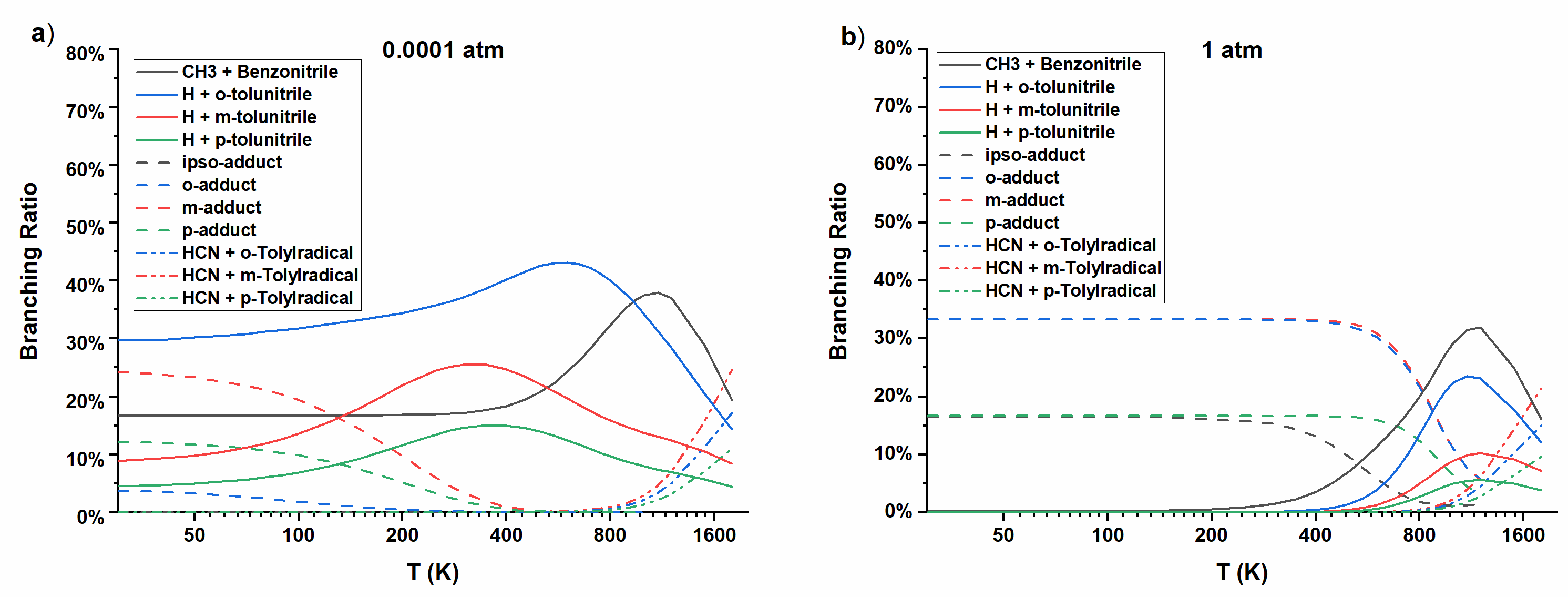}
\caption{Branching ratios of the addition-elimination channels (solid lines), addition channels (dashed line) and abstraction channels (dash-dot-doted lines) for CN + toluene in this work. a) Branching ratios calculated at the pressure of 0.0001 atm; b) Branching ratios calculated at the pressure of 1 atm.
\label{fig:f4}}
\end{figure}

\begin{figure*}[htbp]
\centering
\includegraphics[width=1\linewidth]{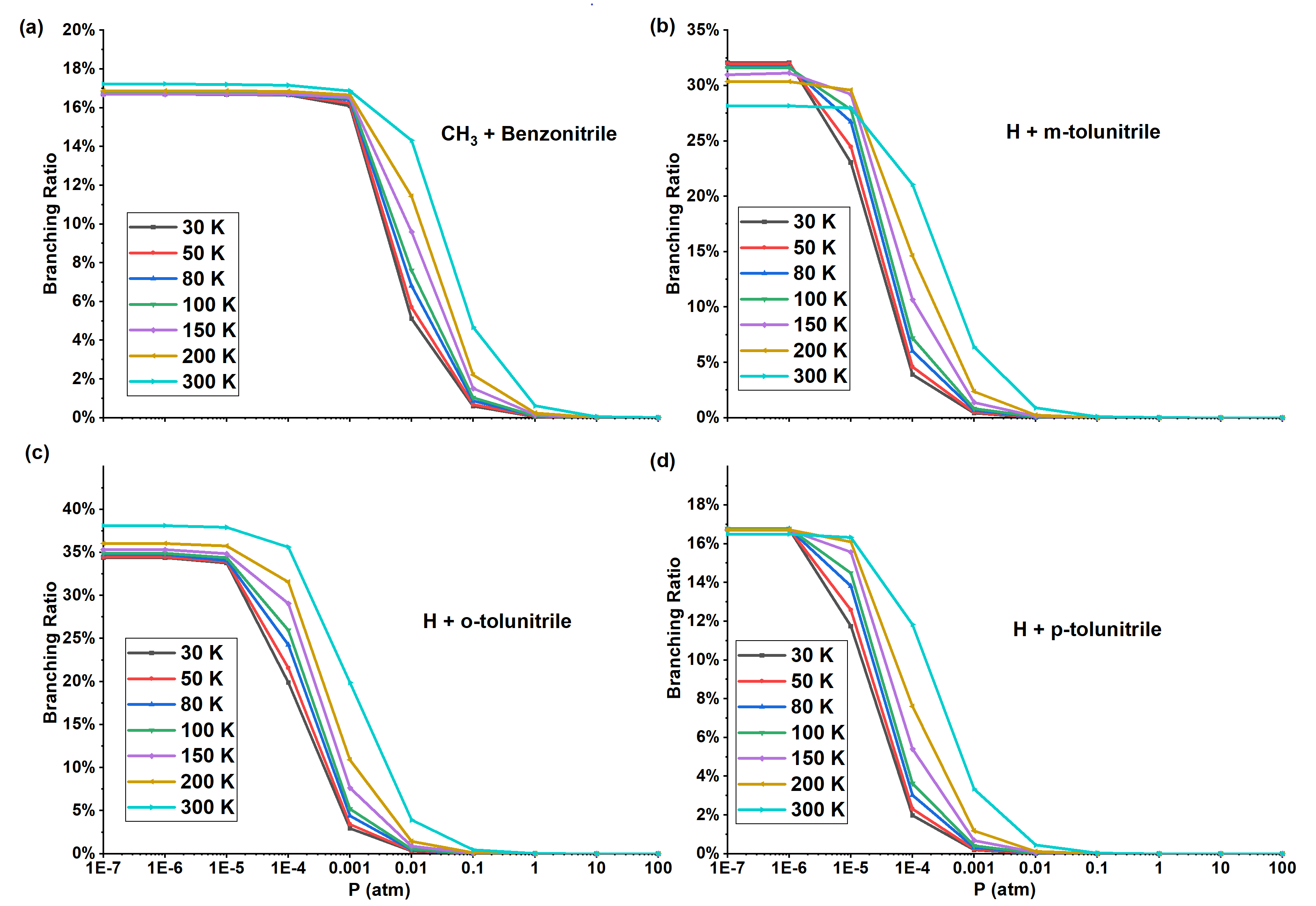}
\caption{Pressure dependence of the branching ratios of various bimolecular products for the CN + toluene reaction at temperatures of 30--300 K and pressures ranging from $10^{-7}$ to 1000 atm with $\rm H_{2}$ as the bath gas. (a): Products of ipso-addition-elimination pathway; (b): Products of o-addition-elimination pathway; (c): Products of m-addition-elimination pathway; (d): Products of p-addition-elimination pathway.\label{fig:f5}}
\end{figure*}

Previous branching ratios information of various channels required by astrochemical models were limited for CN + toluene reaction. The only direct clue of product yields at low temperature comes from the synchrotron VUV-MPIMS measurement at 105 K by Trevitt et al. \citep{20} for the CN + toluene reaction. They concluded the dominant role of the tolunitrile products and a limit of $\leq15\%$ for the $\rm CH_{3}$ elimination channel, while the low signal-to-noise ratios prevent them from obtaining the exact branching ratios of different isomers of tolunitrile. The RRKM/ME calculations here show great advantages for providing the corresponding product yields for various reaction channels at a wide range of conditions \citep{13,24,25,26}. Figure \ref{fig:f4} (a) and (b) shows the temperature dependence of main product yields at low pressure (0.0001 atm) and atmospheric pressure (1 atm) respectively, where the product yeilds of the intermediates are not shown for their minor yields. 
Competition among the addition channels, addition-elimination channels and abstraction channels from different sites proceeds throughout the wide range of temperature and pressure. The yields of various addition-elimination increase initially and then decrease as the temperature rises, while their competitiveness decreases sharply with pressure rising. On the contrary, the stabilization of adducts becomes more favorable at pressures over 0.01 atm, and the yields of adducts decrease monotonously with temperature. The branching ratios for abstraction channels start to compete with the addition-elimination channels at temperatures over ~800 K, where their yields monotonously increase with temperature over a broad range of pressures. Among the three abstraction channels from different sites, the m-abstraction channel edges out the others. Figure \ref{fig:f5} shows the pressure dependence of various bimolecular channels at low and atmospheric temperatures in this work (30--300 K). We concluded that the branching ratios for the addition elimination channels remain virtually unchanged below 250 K and $10^{-5}$ atm. The low-pressure and low-temperature competition results in the dominance of H + o-tolunitrile products with a yield of $\sim 34\%$, followed by the H + m-tolunitrile products with a yield of $32\%$. The $\rm CH_{3}$ + benzonitrile and H + m-tolunitrile products obtain a smaller yield of nearly $ 17\%$ under extreme cold and thin conditions. This study of CN + toluene provides more detailed and complete results of product yields over a wide range of temperatures and pressures. The results show a slightly higher branching ratio of $\rm CH_{3}$ + benzonitrile products compared to the predicted limit from Trevitt et al. \citep{20}.

\section{Astrochemical Implications} \label{sec:implication}

The reaction between toluene and CN radicals shows high efficiency, even under extremely cold conditions, indicating its potential importance in the TMC-1 or Titan’s atmosphere. The astrochemical implications of the detailed and complete product yield information for the CN + toluene reaction is discussed in detail here.

\subsection{TMC-1} \label{sec:TMC-1}

\begin{figure}[htbp]
\centering
\includegraphics[width=0.5\linewidth]{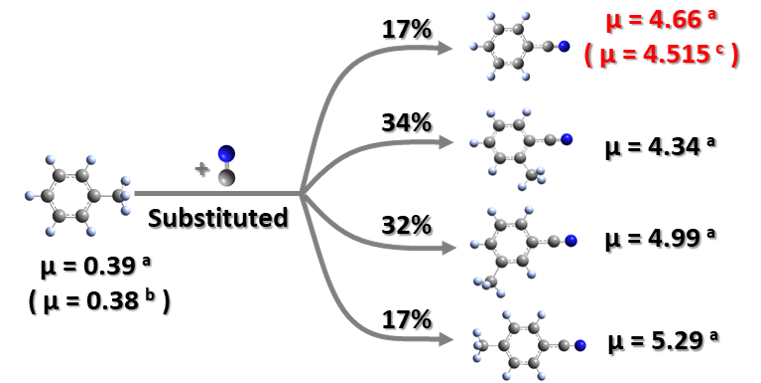}
\caption{Main products and their corresponding branching ratios of CN + toluene reaction under the TMC-1 environment. The dipole moments of the toluene and related products are also demonstrated, where the tags represent different sources of data. $\rm ^{a}$Calculated data at wB97M-V/aug-cc-pVTZ in this work; $\rm ^{b}$experimental data obtained by Rudolph et al. \citep{55}; $\rm ^{c}$experimental data from Wohlfart et al. \citep{56}.  \label{fig:f6}}
\end{figure}

Figure \ref{fig:f6} illustrates the branching ratios and dipole moments of the main products for CN + toluene reaction under the TMC-1 environment (5--10 K, $\rm n_{H}~\thickapprox~10^{4}~cm^{-3}$). Benzonitrile has recently been detected in the molecular cloud TMC-1 by McGuire et al. \citep{5}. The only reaction contributing to the formation of benzonitrile in McGuire et al.’s model of TMC-1 was CN + benzene $\rightarrow$ benzonitrile + H with a rate coefficient of $\rm 3~\times~10^{-10}~cm^{3}~s^{-1}$ \citep{5}. The calculated results in this work show that the CN + toluene reaction may be another candidate for producing benzonitrile in TMC-1 due to its large rate coefficients and considerable branch ratio of nearly 17$\%$ for the benzonitrile + $\rm CH_{3}$ channel \citep{5,53}. The missing CN +toluene reaction in McGuire et al.’s model of TMC-1 may help to explain the underestimation of the benzonitrile column. The results also indicate the important role of the reaction between the $\rm CH_{3}$-substituted aromatic ring and the CN radical during the production of CN-substituted aromatic rings.

This work also reveals the large yields of three isomers of tolunitriles of the CN + toluene reaction as shown in Figure \ref{fig:f6}. All the tolunitriles possess a much higher dipole moment than their parent toluene. Cernicharo et al. \citep{16} reported a high 3$\rm \sigma$ upper limit of $\rm 6~\times~10^{12}~cm^{-2}$ for toluene with the QUIJOTE1 line survey, while the poor dipole moment of toluene hindered the further quantification of its precise column density in the TMC-1. The reaction between toluene and CH radical is an important route of producing styrene in the model of TMC-1. The reaction between styrene and the CH radical will contribute to the production of indene \citep{54}. This work shows that the detection of tolunitriles can serve as a proxy for the identification of toluene in the TMC-1, and the branching ratio information here will constrain the prediction of toluene from the model of the TMC-1.

\subsection{Titan’s Atmosphere} \label{Titan}

Competition between bimolecular and unimolecular products is extremely intense in the PAH/PANH-forming region of Titan’s stratosphere (3–-0.1 mbar; 160–-180 K). Despite the negligible pressure-dependence of total rate coefficients, the yields of different products vary greatly with pressure.
\begin{figure*}[htbp]
\centering
\includegraphics[width=0.8\linewidth]{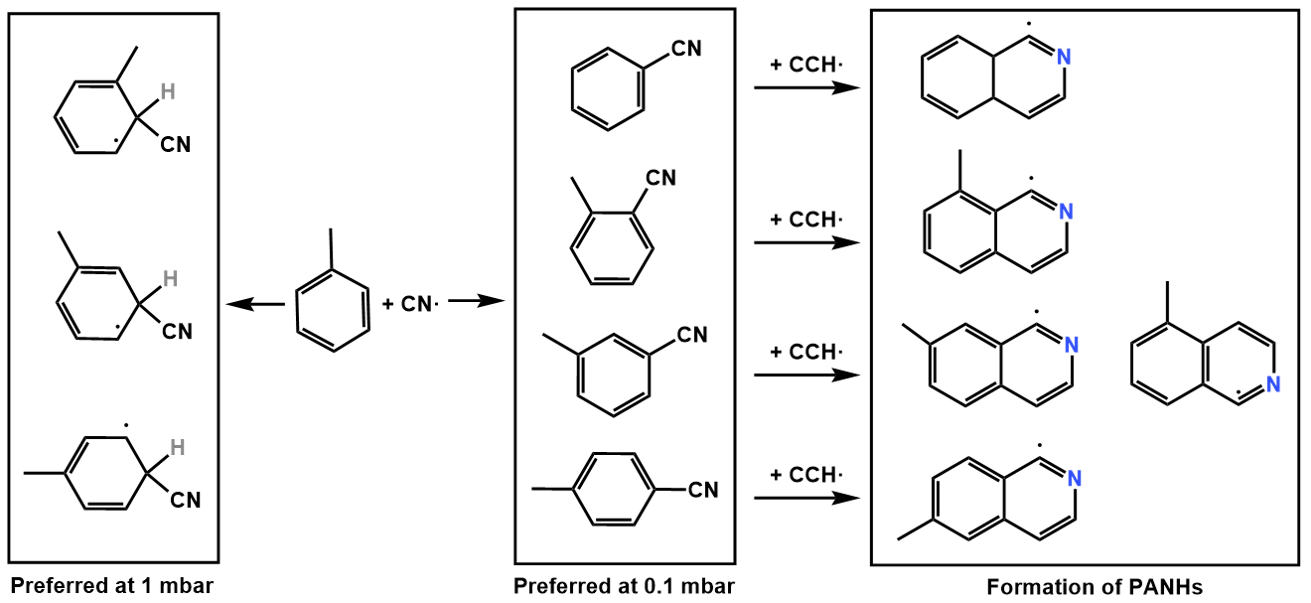}
\caption{Main products of the CN + toluene reaction at Titan’s atmosphere. Potential PANHs produced by the subsequent $\rm C_{2}H$ additions are also shown schematically. Detailed processes and related barriers of the PANH formation are shown in Figure \ref{fig:fA2} in the Appendix.  
\label{fig:f7}}
\end{figure*}
The fractions of tolunitriles, adducts and benzonitrile are 19$\%$--68$\%$, 15$\%$--64$\%$ and 17$\%$ at 150--200 K and 0.0001--0.001 atm, respectively. Landera and Mebel \citep{12,13} show that consecutive $\rm C_{2}H$ additions to aromatic nitriles may contribute to the formation of PANHs. To investigate the similar processes of the “hetero” N atoms doping into the related products of the titled reaction in the PAH/PANH-forming region of Titan’s stratosphere, we determined the reaction pathways of PANH formation from tolunitriles/benzonitrile by the $\rm C_{2}H$ additions, whose details can be found in Figure \ref{fig:fA2} in the Appendix. The $\rm C_{2}H$ addition formed the adducts initially, and the adducts may undergo elimination of the hydrogen atom to form the $\rm C_{2}H$-substituted benzonitrile/tolunitriles or isomerization to form various PANHs shown in Figure \ref{fig:f7}. The submerged barriers for the PANH formation by $\rm C_{2}H$ addition to benzonitrile/tolunitriles indicate that these reactions may be especially efficient in the cold environment of Titan’s atmosphere. However, the byproducts formed by the addition/addition-elimination pathways in Figure \ref{fig:fA2} in the Appendix are also competitive. Further studies are needed to verify their relative yields.

\section{Conclusions} \label{sec:conclusion}

The formation mechanism of aromatic nitriles under cold and collision-free conditions is critical for understanding the tholins’ formation in Titan’s atmosphere and to model the bottom-up formation of PAHs in the ISM. Previous studies of CN + toluene reaction have shown its high efficiency and relationship with benzonitrile formation in the ISM. The exact yields and rate coefficients of the various channels needed in these astrochemical models remain ambiguous. Here, the RRKM/ME calculations reveal the competition of its addition-elimination, addition, and abstraction channels at various temperatures and pressures. In the extreme environment related to the star-forming region of TMC-1, it’s conjectured from our calculations that the fraction of benzonitrile product for the CN reacting with toluene may be approximately 17$\%$, indicating that the CN + toluene reaction can serve as one of gas-phase formation mechanisms for benzonitrile in the models of TMC-1. It’s also revealed  that three isomers of tolunitrile with tremendous yields of the CN + toluene reaction are potential proxies of toluene for the larger dipole moments. In contrast to the negligible pressure-dependence in the awfully cold region of ISM, competition between bimolecular and unimolecular products is extremely intense in the temperature range of Titan’s stratosphere. Under the warmer and denser PANH forming region of Titan’s stratosphere, our results show that the fractions of tolunitriles, adducts, and benzonitrile are 19$\%$--68$\%$, 15$\%$--64$\%$ and 17$\%$ at 150–-200 K and 0.0001–-0.001 atm, respectively. The closed shell products, benzonitrile and tolunitriles, may contribute to the formations of PANHs by consecutive $\rm C_{2}H$ additions. The yield information about aromatic nitriles for the CN + toluene reaction is calculated in this work. Such insights can be applied in future astrochemical models to better understand the evolution of tholins in Titan’s atmosphere and the formation of PAHs in the ISM.

\begin{acknowledgments}
The authors appreciate Prof. Long Zhao of the University of Science and Technology of China for helpful discussions. This study was financially supported by the National Natural Science Foundation of China (22173089, 51876199). Some of the quantum chemical calculations were carried out on the supercomputing system in the Supercomputing Center of the University of Science and Technology of China. We thank LetPub (www.letpub.com) for its linguistic assistance during the preparation of this manuscript.
\end{acknowledgments}

%




\appendix \label{appendix}

\begin{figure}[htbp]
\figurenum{A1}
\centering
\includegraphics[width=0.5\linewidth]{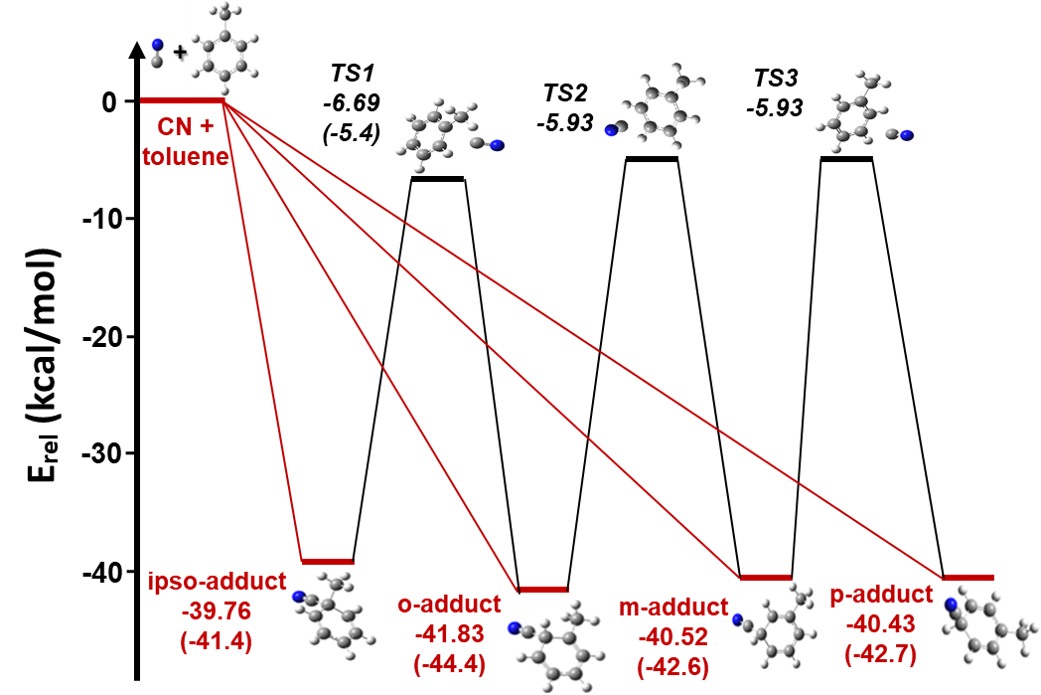}
\caption{Entrance channels of the CN + toluene reaction at the level of CCSD(T)/CBS//M06–2X-D3/6-311+G(d,p). The energetically favorable pathways are represented in red. Energies in parentheses are the results calculated at the level of M06-2X/aug-cc-pVTZ from Messinger et al. \citep{19}.  
\label{fig:fA1}}
\end{figure}

\begin{figure*}[htbp]
\figurenum{A2}
\centering
\includegraphics[width=1\linewidth]{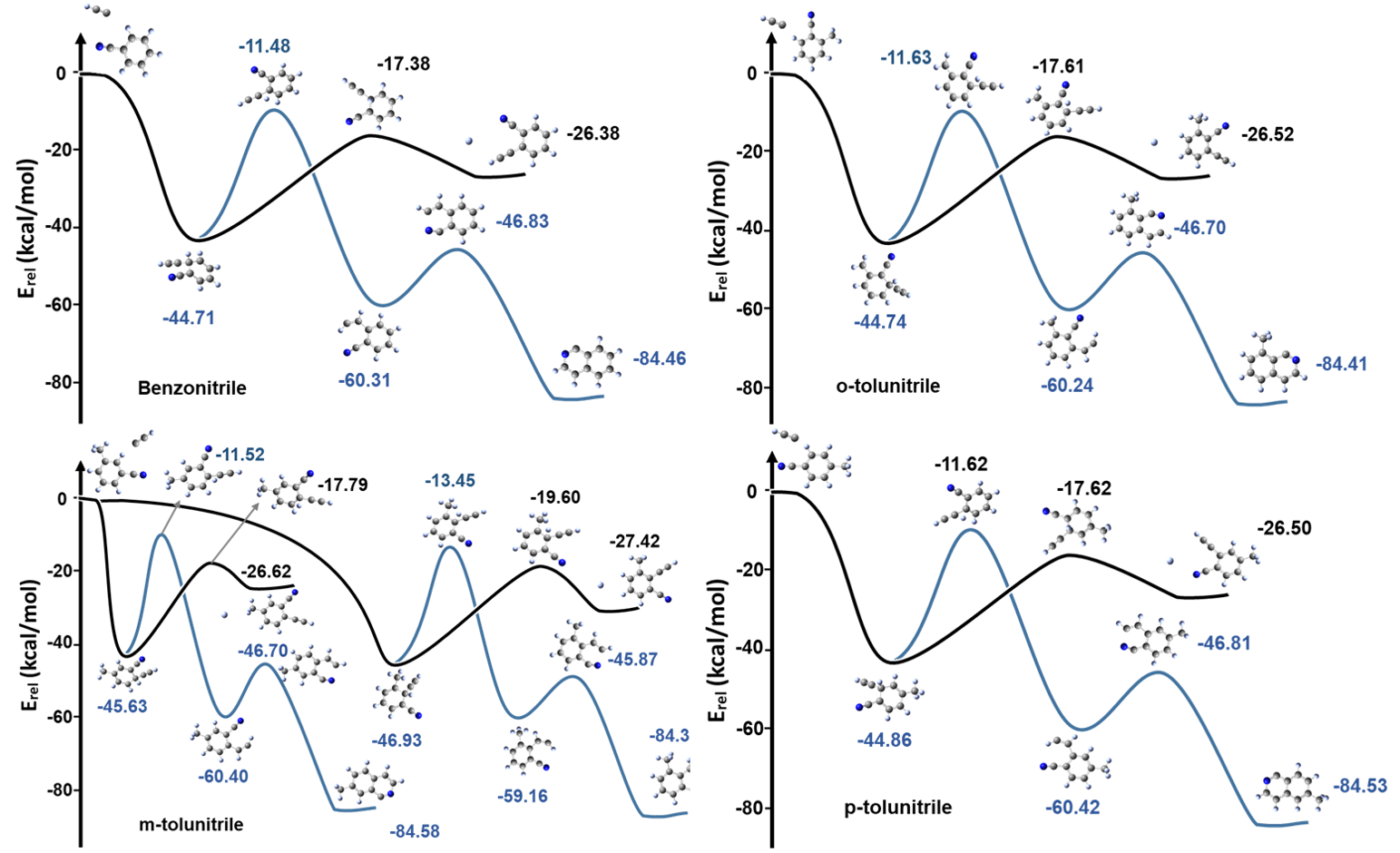}
\caption{The reaction pathways of PANHs formation from tolunitriles/benzonitrile by the $\rm C_{2}H$ additions at the level of M06–2X-D3/6-311+G(d,p). 
\label{fig:fA2}}
\end{figure*}

Table \ref{table:TA1} lists rate coefficients of various channels with $\rm H_{2}$ as the bath gas at 30--200 K, under pressures of $\rm 10^{-7}$, 0.0001, and 0.001 atm. Entrance channels of the CN + toluene reaction in section \ref{PES} are shown in Figure \ref{fig:fA1}. Reaction pathways of PANH formation from tolunitriles/benzonitrile by the $\rm C_{2}H$ addition in section \ref{Titan} are shown in Figure \ref{fig:fA2}.
\startlongtable
\begin{deluxetable*}{cc|ccccccccc}
\tablenum{A1}
\tablecaption{Rate coefficients (in unit of $\rm cm^{3}~s^{-1}$) of main products as a function of the temperature with the bath gas of $\rm H_{2}$ at 30--200 K, $\rm 10^{-7}$, 0.0001, and 0.001 atm. BN and TN is short for benzonitrile and tolunitrile, respectively.}
\label{table:TA1}
\tablehead{
\colhead{T(K)} & \colhead{P(atm)} & \colhead{$\rm CH_{3}$ + BN} & \colhead{H + o-TN} &
\colhead{H + m-TN} & \colhead{H + p-TN} &\colhead{o-adduct} & \colhead{m-adduct} &\colhead{p-adduct}
}
\startdata
{ } &$\rm 10^{-7}$ &7.97$\rm \times 10^{-11}$ &1.64$\rm \times 10^{-10}$ &1.53$\rm \times 10^{-10}$ &7.99$\rm \times 10^{-11}$ &2.51$\rm \times 10^{-16}$ &1.18$\rm \times 10^{-15}$ &5.66$\rm \times 10^{-16}$ \\
30 &0.0001 &7.95$\rm \times 10^{-11}$ &9.48$\rm \times 10^{-11}$ &1.86$\rm \times 10^{-11}$ &9.41$\rm \times 10^{-12}$ &6.42$\rm \times 10^{-11}$ &1.40$\rm \times 10^{-10}$ &7.01$\rm \times 10^{-11}$ \\
{ } &0.001 &7.68$\rm \times 10^{-11}$ &1.41$\rm \times 10^{-11}$ &1.98$\rm \times 10^{-12}$ &9.98$\rm \times 10^{-13}$ &1.45$\rm \times 10^{-10}$ &1.57$\rm \times 10^{-10}$ &7.85$\rm \times 10^{-11}$ \\
\hline
{ } &$\rm 10^{-7}$ &8.67$\rm \times 10^{-11}$ &1.79$\rm \times 10^{-10}$ &1.66$\rm \times 10^{-10}$ &8.70$\rm \times 10^{-11}$ &6.63$\rm \times 10^{-17}$ &3.44$\rm \times 10^{-16}$ &1.64$\rm \times 10^{-16}$ \\
50 &0.0001 &8.66$\rm \times 10^{-11}$ &1.12$\rm \times 10^{-10}$ &2.39$\rm \times 10^{-11}$ &1.20$\rm \times 10^{-11}$ &6.15$\rm \times 10^{-11}$ &1.49$\rm \times 10^{-10}$ &7.45$\rm \times 10^{-11}$ \\
{ } &0.001 &8.43$\rm \times 10^{-11}$ &1.77$\rm \times 10^{-11}$ &2.57$\rm \times 10^{-12}$ &1.29$\rm \times 10^{-12}$ &1.55$\rm \times 10^{-10}$ &1.70$\rm \times 10^{-10}$ &8.52$\rm \times 10^{-11}$ \\
\hline
{ } &$\rm 10^{-7}$ &9.17$\rm \times 10^{-11}$ &1.90$\rm \times 10^{-10}$ &1.75$\rm \times 10^{-10}$ &9.21$\rm \times 10^{-11}$ &1.63$\rm \times 10^{-17}$ &9.47$\rm \times 10^{-17}$ &4.52$\rm \times 10^{-17}$ \\
70 &0.0001 &9.16$\rm \times 10^{-11}$ &1.28$\rm \times 10^{-10}$ &3.02$\rm \times 10^{-11}$ &1.52$\rm \times 10^{-11}$ &5.52$\rm \times 10^{-11}$ &1.52$\rm \times 10^{-10}$ &7.63$\rm \times 10^{-11}$ \\
{ } &0.001 &8.98$\rm \times 10^{-11}$ &2.22$\rm \times 10^{-11}$ &3.33$\rm \times 10^{-12}$ &1.67$\rm \times 10^{-12}$ &1.61$\rm \times 10^{-10}$ &1.80$\rm \times 10^{-10}$ &8.98$\rm \times 10^{-11}$ \\
\hline
{ } &$\rm 10^{-7}$ &9.57$\rm \times 10^{-11}$ &1.99$\rm \times 10^{-10}$ &1.81$\rm \times 10^{-10}$ &9.60$\rm \times 10^{-11}$ &3.81$\rm \times 10^{-18}$ &2.44$\rm \times 10^{-17}$ &1.19$\rm \times 10^{-17}$ \\
90 &0.0001 &9.55$\rm \times 10^{-11}$ &1.43$\rm \times 10^{-10}$ &3.77$\rm \times 10^{-11}$ &1.90$\rm \times 10^{-11}$ &4.79$\rm \times 10^{-11}$ &1.52$\rm \times 10^{-10}$ &7.65$\rm \times 10^{-11}$ \\
{ } &0.001 &9.41$\rm \times 10^{-11}$ &2.73$\rm \times 10^{-11}$ &4.27$\rm \times 10^{-12}$ &2.14$\rm \times 10^{-12}$ &1.63$\rm \times 10^{-10}$ &1.86$\rm \times 10^{-10}$ &9.32$\rm \times 10^{-11}$ \\
\hline
{ } &$\rm 10^{-7}$ &9.74$\rm \times 10^{-11}$ &2.03$\rm \times 10^{-10}$ &1.84$\rm \times 10^{-10}$ &9.77$\rm \times 10^{-11}$ &1.82$\rm \times 10^{-18}$ &1.23$\rm \times 10^{-17}$ &5.99$\rm \times 10^{-18}$ \\
100 &0.0001 &9.72$\rm \times 10^{-11}$ &1.51$\rm \times 10^{-10}$ &4.19$\rm \times 10^{-11}$ &2.11$\rm \times 10^{-11}$ &4.42$\rm \times 10^{-11}$ &1.51$\rm \times 10^{-10}$ &7.61$\rm \times 10^{-11}$ \\
{ } &0.001 &9.60$\rm \times 10^{-11}$ &3.02$\rm \times 10^{-11}$ &4.82$\rm \times 10^{-12}$ &2.41$\rm \times 10^{-12}$ &1.64$\rm \times 10^{-10}$ &1.89$\rm \times 10^{-10}$ &9.47$\rm \times 10^{-11}$ \\
\hline
{ } &$\rm 10^{-7}$ &1.01$\rm \times 10^{-10}$ &2.12$\rm \times 10^{-10}$ &1.89$\rm \times 10^{-10}$ &1.01$\rm \times 10^{-10}$ &2.66$\rm \times 10^{-19}$ &2.28$\rm \times 10^{-18}$ &1.11$\rm \times 10^{-18}$ \\
125 &0.0001 &1.01$\rm \times 10^{-10}$ &1.67$\rm \times 10^{-10}$ &5.34$\rm \times 10^{-11}$ &2.70$\rm \times 10^{-11}$ &3.59$\rm \times 10^{-11}$ &1.46$\rm \times 10^{-10}$ &7.40$\rm \times 10^{-11}$ \\
{ } &0.001 &1.00$\rm \times 10^{-10}$ &3.83$\rm \times 10^{-11}$ &6.49$\rm \times 10^{-12}$ &3.24$\rm \times 10^{-12}$ &1.63$\rm \times 10^{-10}$ &1.95$\rm \times 10^{-10}$ &9.75$\rm \times 10^{-11}$ \\
\hline
{ } &$\rm 10^{-7}$ &1.04$\rm \times 10^{-10}$ &2.20$\rm \times 10^{-10}$ &1.93$\rm \times 10^{-10}$ &1.04$\rm \times 10^{-10}$ &4.79$\rm \times 10^{-20}$ &5.90$\rm \times 10^{-19}$ &2.78$\rm \times 10^{-19}$ \\
150 &0.0001 &1.04$\rm \times 10^{-10}$ &1.81$\rm \times 10^{-10}$ &6.64$\rm \times 10^{-11}$ &3.37$\rm \times 10^{-11}$ &2.87$\rm \times 10^{-11}$ &1.38$\rm \times 10^{-10}$ &7.04$\rm \times 10^{-11}$ \\
{ } &0.001 &1.03$\rm \times 10^{-10}$ &4.77$\rm \times 10^{-11}$ &8.68$\rm \times 10^{-12}$ &4.33$\rm \times 10^{-12}$ &1.60$\rm \times 10^{-10}$ &1.99$\rm \times 10^{-10}$ &9.96$\rm \times 10^{-11}$ \\
\hline
{ } &$\rm 10^{-7}$ &1.07$\rm \times 10^{-10}$ &2.28$\rm \times 10^{-10}$ &1.96$\rm \times 10^{-10}$ &1.07$\rm \times 10^{-10}$ &1.32$\rm \times 10^{-20}$ &5.37$\rm \times 10^{-19}$ &2.48$\rm \times 10^{-19}$ \\
175 &0.0001 &1.07$\rm \times 10^{-10}$ &1.94$\rm \times 10^{-10}$ &8.06$\rm \times 10^{-11}$ &4.13$\rm \times 10^{-11}$ &2.28$\rm \times 10^{-11}$ &1.28$\rm \times 10^{-10}$ &6.56$\rm \times 10^{-11}$ \\
{ } &0.001 &1.06$\rm \times 10^{-10}$ &5.87$\rm \times 10^{-11}$ &1.16$\rm \times 10^{-11}$ &5.77$\rm \times 10^{-12}$ &1.55$\rm \times 10^{-10}$ &2.01$\rm \times 10^{-10}$ &1.01$\rm \times 10^{-10}$ \\
\hline
{ } &$\rm 10^{-7}$ &1.10$\rm \times 10^{-10}$ &2.35$\rm \times 10^{-10}$ &1.98$\rm \times 10^{-10}$ &1.09$\rm \times 10^{-10}$ &7.82$\rm \times 10^{-21}$ &9.18$\rm \times 10^{-19}$ &4.26$\rm \times 10^{-19}$ \\
200 &0.0001 &1.10$\rm \times 10^{-10}$ &2.06$\rm \times 10^{-10}$ &9.54$\rm \times 10^{-11}$ &4.96$\rm \times 10^{-11}$ &1.80$\rm \times 10^{-11}$ &1.15$\rm \times 10^{-10}$ &5.97$\rm \times 10^{-11}$ \\
{ } &0.001 &1.09$\rm \times 10^{-10}$ &7.14$\rm \times 10^{-11}$ &1.55$\rm \times 10^{-11}$ &7.70$\rm \times 10^{-12}$ &1.47$\rm \times 10^{-10}$ &2.01$\rm \times 10^{-10}$ &1.01$\rm \times 10^{-10}$ \\
\enddata
\end{deluxetable*}

 \clearpage






\bibliography{CNtoluene}{}
\bibliographystyle{aasjournal}



\end{document}